\documentclass[%
 reprint,superscriptaddress,
nobibnotes,
 amsmath,amssymb,
 aps,
 prb,
]{revtex4-1}

\usepackage{siunitx}
\usepackage{bm}%
\usepackage{xcolor} 
\usepackage{graphicx}
\usepackage[utf8]{inputenc}
\usepackage{epstopdf}
\usepackage{afterpage}
\usepackage{setspace}
\usepackage{booktabs}
\usepackage[colorlinks=true,linkcolor=blue,citecolor=blue]{hyperref}

\usepackage{mhchem}            
\usepackage[utf8]{inputenc}     
\usepackage[T1]{fontenc}        
\DeclareUnicodeCharacter{0308}{\"}  

\usepackage{upgreek}
\definecolor{customcolor}{RGB}{200, 0, 200}
\expandafter\ifx\csname package@font\endcsname\relax\else
 \expandafter\expandafter
 \expandafter\usepackage
 \expandafter\expandafter
 \expandafter{\csname package@font\endcsname}%
\fi
\begin{document}

\setstretch{1.0}
\title{Machine Learning Potentials for Heterogeneous Catalysis}

\author{Amir Omranpour}
\email{\textcolor{black}{amir.omranpour@rub.de}}
\affiliation{Lehrstuhl f\"ur Theoretische Chemie II, Ruhr-Universit\"at Bochum, 44780 Bochum, Germany}
\affiliation{Research Center Chemical Sciences and Sustainability, Research Alliance Ruhr, 44780 Bochum, Germany}
\author{Jan Elsner}
\affiliation{Lehrstuhl f\"ur Theoretische Chemie II, Ruhr-Universit\"at Bochum, 44780 Bochum, Germany}
\affiliation{Research Center Chemical Sciences and Sustainability, Research Alliance Ruhr, 44780 Bochum, Germany}
\author{K. Nikolas Lausch}
\affiliation{Lehrstuhl f\"ur Theoretische Chemie II, Ruhr-Universit\"at Bochum, 44780 Bochum, Germany}
\affiliation{Research Center Chemical Sciences and Sustainability, Research Alliance Ruhr, 44780 Bochum, Germany}
\author{J\"{o}rg Behler}
\email{\textcolor{black}{joerg.behler@rub.de}}
\affiliation{Lehrstuhl f\"ur Theoretische Chemie II, Ruhr-Universit\"at Bochum, 44780 Bochum, Germany}
\affiliation{Research Center Chemical Sciences and Sustainability, Research Alliance Ruhr, 44780 Bochum, Germany}

\date{\today}

\begin{abstract}

The sustainable production of many bulk chemicals relies on heterogeneous catalysis. The rational design or improvement of the required catalysts critically depends on insights into the underlying mechanisms at the atomic scale. In recent years, substantial progress has been made in applying advanced experimental techniques to complex catalytic reactions \textit{in operando}, but in order to achieve a comprehensive understanding, additional information from computer simulations is indispensable in many cases. 
In particular, \textit{ab initio} molecular dynamics (AIMD) has become an important tool to explicitly address the atomistic level \textit{structure}, \textit{dynamics}, and \textit{reactivity} of interfacial systems, but the high computational costs limit applications to systems consisting of at most a few hundred atoms for simulation times of up to tens of picoseconds. Rapid advances in the development of modern machine learning potentials (MLP) now offer a new approach to bridge this gap, enabling simulations of complex catalytic reactions with \textit{ab initio} accuracy at a small fraction of the computational costs. In this perspective, we provide an overview of the current state of the art of applying MLPs to systems relevant for heterogeneous catalysis along with a discussion of the prospects for the use of MLPs in catalysis science in the years to come.
\end{abstract}

\maketitle 


\section{Introduction}\label{sec:introduction}

Heterogeneous catalysis plays an important role in many industrial processes and environmental applications by enabling chemical reactions of molecular species, which are typically in the gas or liquid phase, at solid catalyst surfaces~\cite{grunert2023catalysis,schlogl2015heterogeneous,fechete2012past}. These catalysts lower activation energies, enhance reaction rates, and offer selective pathways for desired products while substantially reducing energy consumption~\cite{schlogl2015heterogeneous}. Some of many important examples of heterogeneously catalysed processes are, e.g., petrochemical refining~\cite{marcilly2003present}, ammonia synthesis\cite{humphreys2021development}, hydrogenation reactions~\cite{cardenas2013development}, fuel cells~\cite{antolini2007catalysts}, polymerization reactions~\cite{johnson1995new}, and selective oxidations~\cite{schlogl2016selective,najafishirtari2021perspective}.

Atomic-level insights into catalytic processes are essential for advancing the rational design of improved catalysts. 
While experimental techniques for studying catalytic reactions \textit{in operando} have significantly advanced — using methods such as Scanning Tunneling Microscopy (STM)~\cite{gewirth1997electrochemical,itaya1998situ}, Atomic Force Microscopy (AFM)~\cite{gewirth1997electrochemical}, Sum-Frequency Vibrational Spectroscopy (SFVS)~\cite{shen2006sum}, and Surface-Enhanced Raman Spectroscopy (SERS)~\cite{shen2006sum,li2017core} — obtaining comprehensive atomic-scale information from these experiments alone remains challenging~\cite{magnussen2019toward}. Thus, complementary information from theoretical studies is urgently needed. In particular electronic structure calculations, most notably Density Functional Theory (DFT), now enable the study of moderately sized systems, on the order of hundreds of atoms, with good accuracy. Consequently, to date, the most widely used theoretical approaches for investigating catalytic processes at the atomistic level are rooted in a \textit{static} surface science approach~\cite{norskov2009towards,rossmeisl2007electrolysis,hussain2016faraday,skulason2017atomic,norskov2004origin,kulkarni2018understanding}, which involves calculating the thermodynamics of surface reaction intermediates by DFT. While these methods have been very successful in screening and predicting new catalysts, they often rely on rather simple structural models of the system under study, which is sometimes referred to as the complexity gap. For instance, the solvent’s impact is often only captured implicitly or including only a small number of explicit solvent molecules, the dynamic nature of catalytic interfaces is largely ignored, and the synergistic effects of adsorbed species are not explicitly considered. 
Additionally, for reactions at solid-liquid interfaces the structure of a solvent near a surface exhibits significant differences compared to the bulk liquid and, e.g. in case of water, the solvent itself may undergo dissociation and recombination and can thus actively participate in reactions~\cite{creazzo2019dft}.
As heterogeneous catalysis increasingly involves solid-liquid interfaces---owing to their ability to operate under milder conditions and improve selectivity\cite{grunert2023catalysis,najafishirtari2021perspective}---the gap between simplified static theoretical models and real experimental conditions is becoming ever more pronounced.

In parallel to the development of electronic structure theory, various computer simulation techniques, such as Molecular Dynamics (MD)\cite{P2723} and Monte Carlo (MC)\cite{P4339}, have been developed since the 1960s, which are able to account for finite-temperature effects. MD simulations, in particular, have been successfully applied to a range of biological and material systems. To effectively apply these methods---i.e., solving Newton’s equations of motion for systems with many atoms---energies and forces for a large number of structures must be computed. These can be provided in terms of atomistic potentials, which are often called force fields in biology and chemistry\cite{brooks2009charmm,case2008amber,senftle2016reaxff}, or empirical potentials in materials science~\cite{daw1984embedded,baskes1987application,tersoff1988new}. However, most of these models are fitted to a narrow range of experimental or \textit{ab initio} data sets to reproduce specific properties and thus necessarily can only reach a limited accuracy.  This problem is particularly severe for catalytic systems, as not only an accurate representation of multiple bulk phases, i.e., the catalyst and the solvent exhibiting very different bonding patterns and interactions, but also of complex interfaces is required.
 
In 1985, Car and Parrinello proposed a method\cite{car1985unified} to unify MD with electronic structure calculations, laying the foundation for what is now broadly referred to as \textit{ab initio} Molecular Dynamics (AIMD)\cite{marx2009ab}. 
AIMD offers the potential to capture the time evolution of catalytic processes and provide mechanistic insight into reaction pathways with far greater accuracy than is possible using empirical models~\cite{gross2022ab}. Furthermore, an important advantage of AIMD is its general applicability to a wide range of systems of different chemical compositions without the need for any system-specific adjustments. 
However, AIMD comes with significant computational costs, driven by the need to perform electronic structure calculations at each MD time step, which places strong limitations on both the system size and time scale accessible in the simulations.
While enhanced sampling methods such as metadynamics~\cite{laio2002escaping, bussi2020using} and umbrella sampling~\cite{torrie1977nonphysical} can be used to speed up the sampling of rare events that are very relevant in catalytic reactions, the cubic scaling of DFT with the number of atoms ultimately limits the complexity of system that can be handled.

\begin{figure}
    \centering
    \includegraphics[scale=0.45]{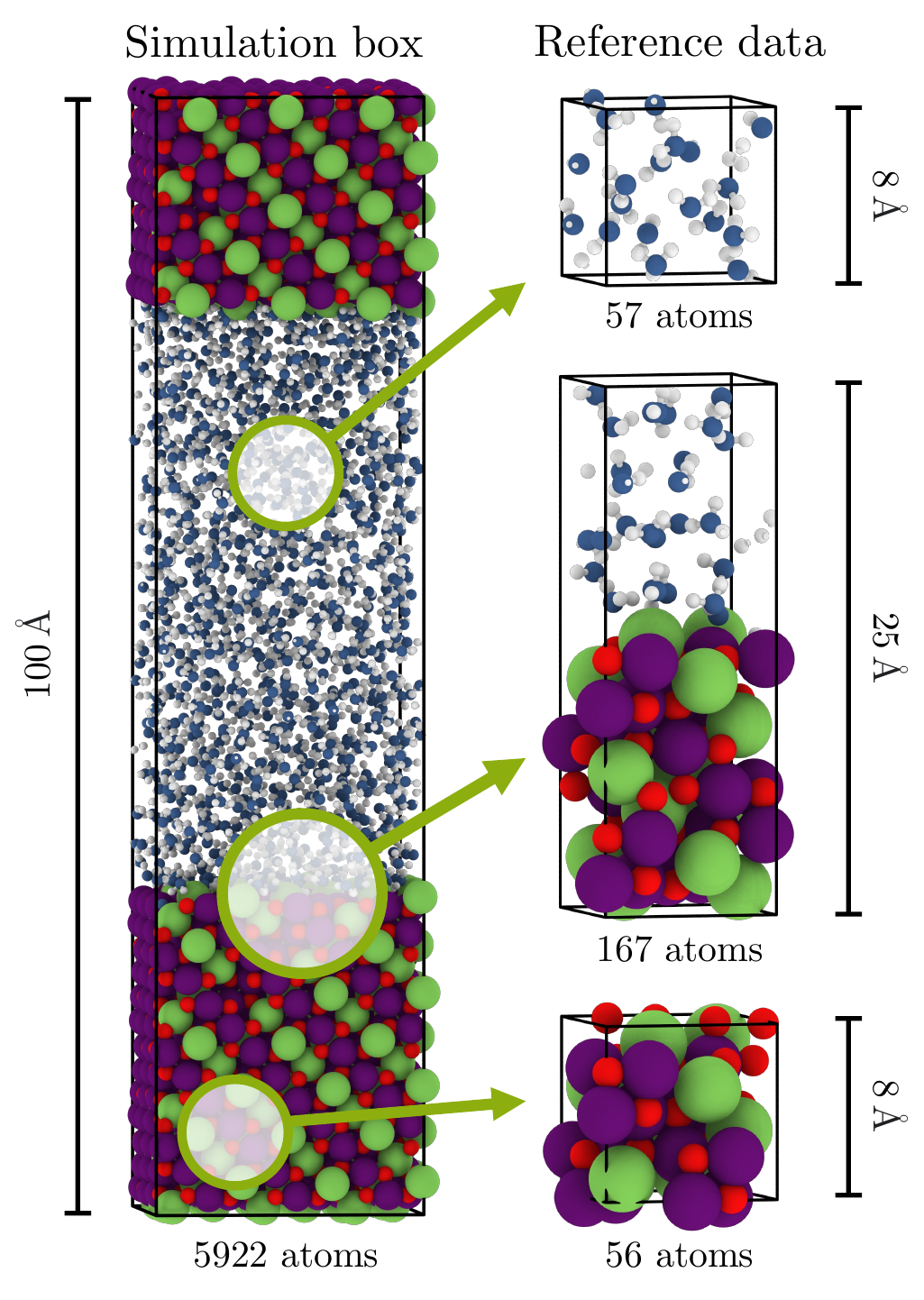}
    \caption{Schematic representation of the LiMn$_2$O$_4$ \{100\}$_\mathrm{Li}$---water interface. On the left, the full simulation box for the final MD simulation is shown. On the right, the smaller reference systems (bulk LiMn$_2$O$_4$, bulk water, and the LiMn$_2$O$_4$ \{100\}$_\mathrm{Li}$---water interface) used for training the MLP model are displayed. Manganese atoms are colored violet, lithium atoms green, oxygen atoms red and hydrogen atoms white. Oxygen atoms of the liquid phase are colored in blue. The figures were created using OVITO Pro (version 3.7.2)\cite{ovito}.}
    \label{fig:MLP-construction}
\end{figure}

Modern machine learning potentials (MLPs) allow to transfer the accuracy of first-principles electronic structure methods to large systems by constructing the high-dimensional potential energy surface based on reference electronic structure calculations for representative atomic configurations using machine learning (ML) techniques~\cite{P4106,P2559,P3033,P4885,P6102,P6112,P6121,P5673}. This makes it feasible to run simulations for systems consisting of thousands of atoms for tens of nanoseconds, which greatly facilitates studying catalytic systems. 
MLPs provide a high numerical consistency with the underlying reference electronic structure method, leading to typical energy errors of around 1~meV/atom and force errors on the order of 100~meV/\AA{}. These errors are much smaller than the uncertainties associated with different exchange-correlation functionals in DFT. Consequently, substituting direct electronic structure calculations by MLPs has only a small impact on the accuracy of the obtained results. Additionally, MLPs are inherently capable of capturing reactive events, i.e., the breaking and formation of chemical bonds, which is crucial for studying catalytic reactions. Figure~\ref{fig:MLP-construction} shows an example for a MLP-driven MD simulation of a reactive LiMn$_2$O$_4$ \{100\}$_\mathrm{Li}$---water interface~\cite{eckhoff2021insights}.

Since the introduction of the first MLP in 1995~\cite{P0316}, MLPs have now become important tools in many fields of chemistry including catalysis, and in fact chemical processes at interfaces have been a driving force in their development. As MLPs now transition from the proof-of-concept phase to mature simulation tools, this article aims to provide a perspective on their growing impact on heterogeneous catalysis research. 
As heterogeneous catalysis is a very broad topic, it is impossible to cover all aspects of using machine learning in this field. In this perspective, we will thus focus on its use for gaining atomic-level insights, and more specifically for representing potential energy surfaces governing catalytic reactions, which can then be used in large-scale simulations. Still, the examples discussed here can only cover some of the systems published in the literature, and other aspects and systems can be found in several related publications~\cite{chen2023machine,tangmachine,cheng2024leveraging,choung2024rise,zhou2023machine,houimproving,P5954,ma2020machine,mou2023bridging, P6771}. 
We will not cover the more general application of machine learning techniques in heterogeneous catalysis, such as using ML to predict new catalysts bypassing the atomic level or exploring catalytic reaction networks, which are equally active fields of research~\cite{chen2022universal,margraf2023exploring}. 

This perspective is organised as follows: first, a concise overview of MLPs is provided, followed by a summary of the achievements of MLPs in elucidating different types of processes in heterogeneous catalysis. Next, several key considerations and challenges in using MLPs in catalysis are addressed, along with potential solutions to be developed. Finally, an outlook on possible future directions of the field is given. 

\section{Machine Learning Potentials}\label{sec:methods}

In recent years, the development of atomistic potentials using machine learning has become a very important topic and many different methods have been proposed, each of which has its own advantages and disadvantages. In this perspective focusing on the application of MLPs in heterogeneous catalysis we do not attempt to provide a comprehensive overview of all the methodical details of MLPs, and we will just provide a bird's eye view of the field. Instead, the interested readers are referred to several dedicated reviews addressing all the aspects of MLPs and their training in great detail~\cite{P2559,P3033,P4885,P6102,P6112,P6121,P5673,P5788,P5673,P5793,P4263,P6018,P5971,P6131,P5977,P6631,P6548}.

\textbf{\textit{Physical Classification of MLPs}} There are several ways to classify the many different flavors of MLPs, each focusing on different aspects. Two possible viewpoints for such categorizations are centered either on the physics governing the atomic interactions~\cite{P5977,P6018} or take a mathematical perspective~\cite{P6121}. The former, which is of particular interest for catalysis, is based on the nature and range of atomic interactions that can be described, and introduces four different generations of MLPs. 

The first generation of MLPs has been restricted to very low-dimensional systems, e.g., diatomic molecules in the gas phase approaching surface models, in which the atomic positions of the surface had to be frozen to reduce the complexity of the potential energy landscape. This limitation was overcome by Behler and Parrinello with the development of high-dimensional neural network potentials (HDNNP) in 2007~\cite{P1174}. HDNNPs introduced a decomposition of the total energy of the system into a sum of local environment-dependent atomic energy contributions $E_i$,
\begin{equation}
E = \sum_{i=1}^{N_{\mathrm{atom}}} E_i,
\end{equation}
where $N_{\rm atom}$ is the total number of atoms in the system. Each atomic energy is then provided by machine learning, e.g., by an atomic neural network. This approach made it feasible to construct MLPs for condensed-phase systems containing large numbers of atoms such as solid surfaces central to heterogeneous catalysis. However, due to the locality approximation, interactions beyond the cutoff—typically in the range of 5 to 10 \AA{}—are only included in an averaged manner. This class of strictly local MLPs is nowadays categorized as second-generation (2G), and many different 2G potentials are now available, which provide very accurate energy surfaces~\cite{P1174,P2630,drautz2019atomic,P4945,shapeev2016moment,P5596}. 
However, for some systems it may be important to incorporate long-range interactions beyond the cutoff or even global phenomena like nonlocal charge transfer, for instance in some complex oxide surfaces.
Overcoming these limitations is the aim of third- and fourth-generation MLPs.

In third-generation MLPs, electrostatic interactions are modeled using environment-dependent charges represented by machine learning models~\cite{P2391,P2962,P3132,P5577,P5313,P5885,P5372,P6200,P5205}. These charges are used to compute the electrostatic energy, which is then combined with the short-range component given by Eq.~(1) to yield the system’s total energy. By training the short-range component to capture only the non-electrostatic contribution to the total energy, double-counting of energy contributions is avoided. 
The remaining shortcomings of third-generation MLPs stem from the assumed locality of atomic charges, which makes it impossible for such models to capture long-range charge transfer that may occur in certain systems~\cite{P5977}. 
Several fourth-generation MLPs have been proposed~\cite{P4419,P5859,ko_fourth-generation_2021,khajehpasha2022cent2,unke_spookynet_2021}, which make use of global information to include these effects. 

Apart from these different capabilities of MLPs to incorporate explicit physics, all MLPs inherit the intrinsic limitations of the underlying training data, in particular, the accuracy of the employed electronic structure level of theory. For instance, if certain interactions, e.g., dispersion interactions, are poorly described in DFT calculations, this will carry over to the constructed MLP. A more detailed discussion of the challenges and prospects of using MLPs to study heterogeneous catalysis will be presented in Section~\ref{sec:discussion}.

\textbf{\textit{Classification by Representation and Learning}} 
From the mathematical viewpoint, 
MLPs can be classified based on how they handle the two central tasks~\cite{P6121}:
(a) \textit{representing} the structure of the system and (b) \textit{learning} the relationship between the representation and the associated potential energy surface (PES). According to this classification, MLPs can be divided into three families:

(1) One family uses predefined descriptors in combination with nonlinear regression. This family typically employs descriptors that respect the three mandatory invariances of the atomic environment—translational, rotational, and permutational symmetries. Most of these descriptors are strictly local, defined by a specific cutoff radius. The relationship between the atomic environment descriptors and the associated PES is then learned using nonlinear fitting functions through either shallow or deep neural networks or kernel regression methods. Some typical representatives of this family include second-generation MLPs like HDNNPs~\cite{P1174}, Gaussian Approximation Potentials (GAP)~\cite{P2630}, and Deep Potential Molecular Dynamics (DeePMD)~\cite{P5596}.

(2) Another family uses 
symmetric basis functions in combination with linear regression, 
albeit with non-linearity included in the basis functions. Some typical MLPs in this category are Moment Tensor Potentials (MTP)~\cite{shapeev2016moment} and Atomic Cluster Expansion (ACE)~\cite{drautz2019atomic}.

(3) More recently, a third family has been introduced using message-passing neural networks (MPNN)~\cite{P5368}. In this approach, both the representation and learning tasks are addressed simultaneously. Here, the descriptors are \textit{learned} by the neural network during training. Some examples in this category include DTNN~\cite{schutt2017quantum}, SchNet~\cite{P5366}, NequIP~\cite{batzner20223}, Allegro~\cite{musaelian2023learning}, and MACE~\cite{P6572}.

In spite of the large number of MLPs that have been published to date,
they have not been equally used 
in studies relating to heterogeneous catalysis. 
A review of the literature indicates that, to date, HDNNP and DeePMD are by far the most commonly applied methods in this context. 
However, several types of MLP, including MPNNs, have been suggested only recently, therefore it is expected that the number of studies employing these newer methods will grow substantially in the coming years. 

\section{Applications to Heterogeneous Catalysis}\label{sec:applications}

 \subsection{Clusters and Surfaces}\label{subsec:other}

\textbf{\textit{Clusters}} 
Clusters are important in catalysis, where they are commonly utilized as nanoparticles or nanoclusters to maximize the active surface area for chemical reactions. A key feature of these nanoparticles is their microstructure and morphology, which can be customized to enhance both functional properties and stability\cite{ghosh2012core}. Many catalysts, in fact, consist of metal clusters supported on oxides. The first applications of MLPs for metal and oxide clusters were reported for copper~\cite{P3114} and zinc oxide clusters~\cite{P2962} using HDNNPs. Further HDNNPs have been reported for Cu–Au nanoalloys employed in grand canonical MD simulations\cite{P4474}. For brass (Cu-Zn), HDNNP studies\cite{P5851,P6179} have shown that the element distribution within the nanoparticles is inhomogeneous, with zinc concentrated in the outermost layer and copper enriched in the subsurface layer. Alloying within the nanoparticle core occurs only at high zinc concentrations, forming crystalline bulk $\alpha$-brass patterns. 
The nanoparticles' melting temperature decreases with higher zinc content, consistent with the bulk phase behavior of brass.
In Pt-Rh alloys, studies using ACE\cite{P6513} have shown that Pt atoms segregate on the surface, forming a monolayer, which contrasts with experimental core-shell nanoclusters having thicker Pt shells that are stabilized kinetically, not thermodynamically. For Au nanoparticles, HDNNP-driven simulations\cite{P4967} revealed a rigid outer atomic layer compared to the core, where changes in surface atom coordination at around 300~K result in surface defects. The simulations also revealed a dynamic coexistence of solid-like and liquid-like phases near the melting transition. Other studies on Au~\cite{P4546,P6042}, Pt~\cite{P5959}, and Na clusters~\cite{P4544} using HDNNPs further confirm the reliability and robustness of MLP-based simulations. The dissociation of a CO$_2$ molecule on Cu nanoclusters was investigated using DeePMD simulations
and well-tempered metadynamics\cite{gong2024machine}, showing that the nanoclusters exhibit surface pre-melting behavior which significantly affects catalytic activity. Point defects, such as vacancies and ad-atoms were found to reduce the surface melting temperature, enabling reactions to occur under milder conditions.

\textbf{\textit{Solid Surfaces}} 
The atomic environments on metal surfaces significantly differ from those in the bulk. Notably, reconstructions or imperfections on real surfaces can result in highly complex atomic configurations. Using ZnO as a case study of a multicomponent system~\cite{P2962}, the third-generation HDNNP was introduced by incorporating environment-dependent charges to account for long-range electrostatic interactions. Copper~\cite{P3114} was also investigated, covering both bulk and various surface structures, including defects. The neural network potential accurately reproduced key properties such as lattice constants, cohesive energies, and surface energies.
The oxidation of flat and stepped Pt surfaces, was studied using  embedded atom neural network (EANN) potentials combined with grand-canonical Monte Carlo simulations~\cite{xu2022atomistic}, revealed the formation mechanism of the square planar \(\text{PtO}_4\) oxide unit on a flat Pt surface. 


\textbf{\textit{Clusters at Surfaces}}
Studies of copper clusters supported on zinc oxide and their extension to the ternary CuZnO system using HDNNPs~\cite{P3827, P5899} revealed a variety of structural patterns. 
Exploration of copper-ceria (CuO/CeO$_2$) catalysts for low-temperature CO oxidation revealed that the surface-substituted Cu$_y$Ce$_{1-y}$O$_{2-x}$ phase is more catalytically active than the bulk CuO phase~\cite{P4475}. This was supported by \textit{in situ} X-ray absorption spectroscopy, electron microscopy, and MLP-driven simulations, which revealed copper ion segregation to the ${100}$ surfaces of nanoparticles.
Additionally, the dynamic restructuring of 2~nm Pt nanoparticles on SiO$_2$ in reactive environments was studied using \textit{in situ} spectroscopy and MD simulations with Allegro~\cite{owen2024surface}, revealing that nanoparticle surfaces lose their atomic order when exposed to CO gas, while their cores remain bulk-like. This challenges traditional models that assume idealized faceting, highlighting the need for models that account for realistic surface structures to predict catalyst function and stability.

\subsection{Solid-Gas Interfaces}\label{subsec:gassurface}

\textbf{\textit{Early MLPs}}
Since the early days of MLPs, their development was motivated by the challenge of accurately representing the PES for gas-surface dynamics. Before the breakthrough of MLPs, 
extensive work during the 1990s and early 2000s focused on analytic PESs for gas-surface dynamics, employing many different approaches~\cite{P0221,P0222,P0234,P0507,P0449,P2690}. A primary limitation of these analytic PESs was their restriction to six dimensions for diatomic molecules on surfaces. Additionally, simulating chemical reaction probabilities, particularly in gas-surface dynamics, demands a precise mapping of energy landscapes. Even minor errors in reaction barriers can lead to significant deviations in predicted reaction probabilities, posing an additional challenge for PES representation.

In response to these limitations, Blank et al.~\cite{P0316} introduced the use of neural networks as a general, nonlinear fitting approach that avoids assumptions about the potential energy surface topology. This approach, using a limited set of data points, demonstrated the ability to model complex chemical interactions with high accuracy. It showed, for the first time that feed-forward neural networks can accurately model PESs, outperforming traditional methods like splines. This technique was applied to systems such as CO on Ni(111) and H$_2$ on Si(100), providing precise predictions of the potential energy.

In 2004, Lorenz et al.~\cite{P0421} took the next step by further reducing the number of training points by exploiting the underlying surface symmetries in periodic slabs. They demonstrated the accuracy and efficiency of neural network potentials (NNPs) for H$_2$ interacting with the (2$\times$2) potassium-covered Pd(100) surface. The sticking probability of H$_2$/K(2$\times$2)/Pd(100) was determined by MD simulations on the neural network PES and compared to results obtained using an independent analytical interpolation. It was shown that, by accounting for the symmetries underlying the particular system and incorporating feedback from dynamical simulations, a relatively moderate number of training points is needed to obtain a reliable fit~\cite{P0620}. 

Similar methodologies were employed by Behler et al. to study O$_2$ dissociation on Al surfaces~\cite{P0662}. Symmetrized functions were introduced~\cite{P1388} to fully account for the surface's symmetry and nonadiabatic effects were also investigated~\cite{P1820}. Instead of molecular coordinates, the symmetrized functions, systematically constructed from atomic Fourier terms, were used as inputs to the neural network. This approach was validated for O$_2$ interacting with the Al(111) surface. Building on this work, additional studies for O$_2$ dissociation on Al were conducted by Carbogno et al.~\cite{P1996, P2536}.

O$_2$ adsorption on Ag surfaces was also investigated~\cite{P3371} and it was shown that the dissociation probabilities match experimental data, suggesting that the surface’s inertness is primarily due to energy barriers, with spin or charge nonadiabaticity playing a negligible role. Later, the conventional view that physisorption states significantly influence molecular scattering experiments was challenged~\cite{P4022}. Despite the inability of semilocal DFT to accurately capture long-range van der Waals interactions and the absence of physisorption wells, experimental scattering trends were successfully reproduced in the simulations. It was suggested that molecular scattering is more influenced by the repulsive walls associated with chemisorption rather than by physisorption, casting doubt on the use of scattering data as indirect evidence for the existence of physisorption states.

\textbf{\textit{Ammonia Decomposition}}
The work discussed so far represents initial efforts employing first-generation MLPs. Since then, MLPs have reached a level of maturity that allows them to be used to study some of the key challenging systems in catalysis, such as the Haber-Bosch process for ammonia synthesis\cite{humphreys2021development}, which is essential for producing fertilizers and sustaining global food production. Traditionally, theoretical modeling at \textit{operando} industrial-level temperatures was unfeasible, often relying on extrapolations from lower-temperature results. However, the validity of this approach has been recently questioned by a series of works by Parrinello et al.~\cite{bonati2023role,purcel2024iron,perego2024dynamics,tripathi2024poisoning}. In particular, these studies directly highlight the role of dynamics in ammonia decomposition on the iron surface relevant to the Haber-Bosch process.

In the Haber-Bosch process, the rate-limiting step is believed to be the N$_2$ decomposition on Fe catalysts, due to the strong triple bond of N$_2$ molecules. N$_2$ decomposition on Fe(111) was investigated~\cite{bonati2023role} using DeePMD and metadynamics simulations, revealing the disruptive effect of dynamic changes on the Fe(111) surface morphology, which significantly influences nitrogen adsorption and dissociation, particularly at elevated temperatures. The study highlights the risk of extrapolating low-temperature results to \textit{operando} conditions (700 K for the N$_2$ decomposition), due to the nonlinearity of catalytic behavior with temperature, and demonstrates that catalytic activity can only be accurately inferred from simulations that explicitly account for dynamics.
Additionally, it has been shown~\cite{purcel2024iron} that ammonia decomposition over a wustite-based bulk iron catalyst results in the formation of iron nitrides (Fe$_4$N and Fe$_2$N) at lower temperatures. The decomposition of Fe$_4$N into Fe and N$_2$ was identified as the rate-determining step, with activation energies of 172 and 173 kJ/mol, respectively. MD simulations demonstrated that nitrogen migration into the bulk of the catalyst is favored over recombination at the surface, significantly impacting the efficiency of nitrogen desorption and nitride formation.

\textbf{\textit{Al$_2$O$_3$(0001) Surface}}
The Al$_2$O$_3$(0001) surface is important in catalysis as it serves as a stable support material for dispersing active catalytic species, enhancing activity and durability. Its well-defined atomic structure and low reactivity make it suitable for high-temperature catalytic processes, while its surface properties allow for controlled adsorption and surface reactions, making it valuable for studying heterogeneous catalysis. The stoichiometric reconstruction of the Al$_2$O$_3$(0001) surface---considered one of surface science's mysteries~\cite{chame1997three}---was recently investigated~\cite{hutner2024stoichiometric} using noncontact atomic force microscopy (nc-AFM) and DFT calculations enhanced by VASP's on-the-fly MLPs~\cite{jinnouchi2019fly,jinnouchi2019phase,jinnouchi2020descriptors}, based on GAP. Imaging revealed the lateral atomic positions, while theoretical analysis indicated that aluminum rehybridization enables bonding with subsurface oxygen atoms, significantly stabilizing the reconstruction. In another study using an HDNNP, hydrogen atom scattering at the Al$_2$O$_3$(0001) surface was examined~\cite{P6672}. The best agreement between experimental atom beam scattering and theory occurs at large initial kinetic energies and at both very low and high scattering angles, attributed to scattering from top-layer aluminum atoms. In contrast, lower initial kinetic energies result in greater kinetic energy loss in the MD trajectories compared to experiment. Additionally, scattering at oxygen sites generally leads to larger discrepancies.

\textbf{\textit{Other Gas-Surface Systems}}
Numerous other studies have employed MLPs to model gas-surface dynamics. For instance, NNPs have been employed to investigate H$_2$ dissociation on Pt(111) and Cu(111), demonstrating that reduced energetic corrugation broadens the reaction probability curve~\cite{P1786}. In a combined HDNNP and STM study~\cite{sumaria2022atomic}, it was found that high CO coverage at a Pt step edge induces the formation of atomic protrusions composed of low-coordination Pt atoms. These atoms then detach from the step edge, forming sub-nano-islands on the terraces, where the CO adsorbates stabilize the under-coordinated sites. H$_2$ dissociation on Pt(111) was also examined using an on-the-fly trained Sparse Gaussian Process (SGP) potential~\cite{vandermause2022active}, while H$_2$ dissociation on curved Pt surfaces was studied with HDNNPs~\cite{gerrits2021accurate}.

Investigation of N$_2$O dissociation on Cu(100)~\cite{P3152} showed that  N\(_2\)O initially weakly adsorbs on Cu(100) before reaching a stable chemisorbed state at a hollow site, with dissociation into N\(_2\) and adsorbed oxygen becoming favorable at higher translational energies. Vibrational shifts in adsorbed N\(_2\)O reflect bond weakening, facilitating dissociation, while changes in rotational temperature have minimal effect on this process. The quantum mechanics/molecular mechanics (QM/MM) embedding method for O$_2$ on Pd(100), where a NNP was used to surrogate DFT calculations, revealed significant energy release during dissociation. This release resulted in highly mobile oxygen atoms, challenging the assumption of instantaneous thermalization in catalytic processes~\cite{P4979}. 
NNP-based studies of HCl on Au(111)~\cite{P5764} showed that the RPBE functional produces higher reaction barriers and lower probabilities than PBE, improving agreement with the experiment, while reactivity is highly sensitive to the rovibrational state population. Surface atom motion, nonadiabatic effects, and moderate charge transfer at the transition state influence the reaction only modestly.
A 15-dimensional PES for CH$_4$ on Ni(111) was found to accurately reproduced methane dissociation dynamics, providing valuable insights for industrial applications~\cite{P5334}. Additionally, the use of the Allegro architecture~\cite{musaelian2023learning} combined with enhanced sampling techniques revealed a dynamic interplay between CH$_4$ and the Ni catalyst, highlighting the increased mobility of adsorbed species, especially at higher temperatures~\cite{xu2024molecular}.

The interactions of N$_2$ with Ru(0001) were explored using an HDNNP trained on RPBE data, showing agreement with experimental molecular beam sticking probabilities~\cite{P5069}. Further analysis revealed that vibrational excitation is more efficient than translational energy in overcoming activation barriers~\cite{P5402}. HDNNPs were also able to provide accurate reaction probabilities for the highly activated reaction of CHD$_3$ on Cu(111)~\cite{P5797}, the effect of orbital-dependent electronic friction on the description of reactive scattering of N$_2$ from Ru(0001)~\cite{P5763}, and H-atom on free-standing graphene~\cite{P5868}. For oxygen on Pd surfaces, HDNNP models were able to predict adsorption energies and diffusion barriers accurately, though improvements are needed at high coverages due to surface reconstructions~\cite{P5112}. Hydrogen scattering on copper surfaces has also been evaluated using various MLP models.~\cite{stark2024benchmarking,P6531}. 
A study of CO on Ru(0001) using EANN potentials~\cite{zugec2024understanding} offered strong support for describing both photoinduced desorption and CO oxidation through nonequilibrated, yet thermal, hot electrons and phonons. The MACE potential was applied to investigate ammonia decomposition on iron-cobalt alloy surfaces~\cite{perego2024data}.

\subsection{Solid-Liquid Interfaces}\label{subsec:solidliquid}

Solid-liquid interfaces are of constantly increasing importance in heterogeneous catalysis and electrocatalysis. 
Even pristine surfaces in contact with pure water display significant complexity due to the active role of interfacial water molecules, which may undergo dissociation and recombination — particularly on transition metal oxides. An accurate description of the interface therefore requires an explicit description of water that also captures reactivity. 

Early work on solid-liquid interfaces using HDNNPs
investigated the effect of solvation on the surface composition of Au-Cu nanoparticles\cite{artrith2014understanding}. When solvent water molecules were included, a mixed Au-Cu surface was preferred, while a core-shell structure was predicted in a vacuum. Soon after, 
the question of how much water is needed to achieve bulk-water-like behavior at various prototypical Cu-water interface models was addressed using an HDNNP\cite{natarajan2016neural}. It was found that a water film thickness of at least 40 \text{\AA} is required to prevent artificial interactions between opposing surfaces and to ensure bulk-like properties in the central region of the water film — a scale that is computationally prohibitive for ab initio models.

\begin{figure}
    \centering
    \includegraphics[scale=0.4]{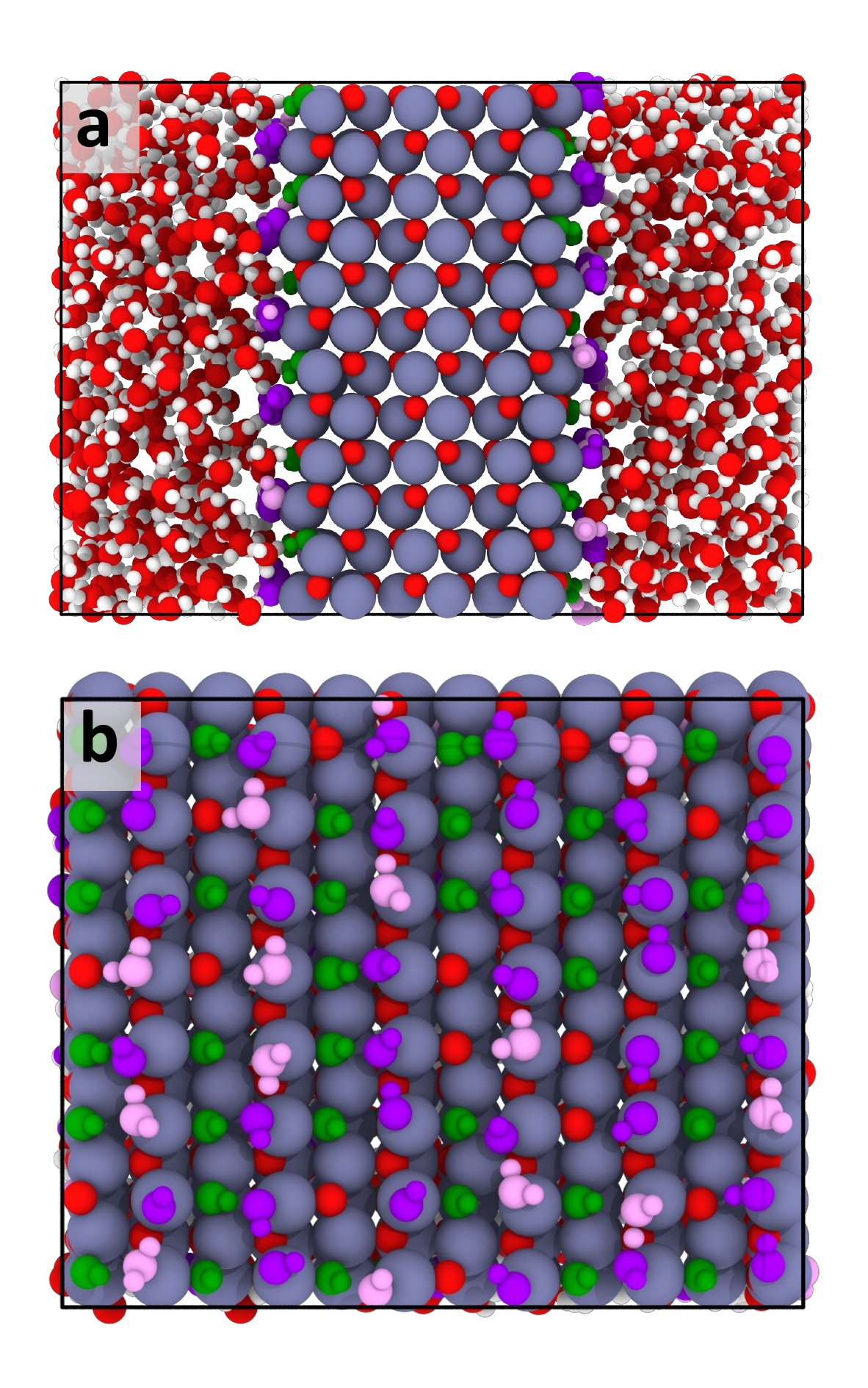}
    \caption{Snapshots of the ZnO(10$\overline{1}$0)-water interface, illustrating the presence of dissociated water at the surface. (a) Side view of the interface model. (b) Top view displaying only the first layer of adsorbed and dissociated water. Adsorbed water molecules, adsorbed hydroxide ions and surface hydroxyl groups are colored in light pink, purple and green, respectively. 
    Hydrogen atoms are assigned to their nearest oxygen atom, while adsorbed species are defined as those with oxygen atoms within 2.5~\text{\AA} of a surface zinc atom.}
    \label{fig:ZnO_water_interface}
\end{figure}

Much attention has been given to the extent of water dissociation and proton transfer (PT) mechanisms at oxide-water interfaces. 
The first MLP-based studies of this kind focused on ZnO-water interfaces\cite{quaranta2017proton, quaranta2018maximally, quaranta2018structure, hellstrom2019one}. 
Simulations of the ZnO(10$\overline{1}0$)\cite{quaranta2017proton} and ZnO(11$\overline{2}$0)-water interfaces\cite{quaranta2018structure} using an HDNNP revealed two dominant types of interfacial PT reactions: PT between surface oxygen atoms and adsorbed hydroxide ions (surface-PT), and PT between adsorbed water molecules and neighbouring adsorbed hydroxide ions (adlayer-PT). Notably, fluctuations in the local hydrogen-bonding environment were found to have a significant effect on PT barriers and corresponding rates\cite{quaranta2017proton}. 
The influence of surface proximity and local hydrogen bonding network on the anharmonic OH stretching vibrations of water and hydroxide ions near the interface were also investigated at the ZnO(10$\overline{1}0$)-water interface\cite{quaranta2018maximally}.
A representative snapshot of this system is shown in Figure~\ref{fig:ZnO_water_interface}, where approximately 70~\% of surface oxygen sites are hydroxylated with protons arising from dissociated water\cite{quaranta2017proton}. 

Consecutive proton transfer reactions at the interface can lead to long-range Grotthuss-like proton diffusion. This was investigated at the ZnO(10$\overline{1}$0) and (11$\overline{2}$0) facets, which showed very different behaviour: proton-diffusion on ZnO(10$\overline{1}$0) was found to be quasi-one-dimensional, whereas on ZnO(11$\overline{2}$0) two-dimension proton transfer was observed\cite{hellstrom2019one}. These differences highlight the strong influence of surface morphology on proton transport pathways. 
Similar observations have been reported in a recent study on long-range proton transport at the CeO$_2$(111) and (110)-water interfaces using DeePMD simulations, which demonstrated significantly more active proton transport on the (111) facet compared to (110)\cite{kobayashi2024long}.

MLPs have been used extensively to study the water interfaces of TiO$_2$\cite{andrade2020free, schran2021machine, o2023elucidating, li2023thermal, zhuang2022resolving,wen2023water, zeng2023mechanistic}, an important system for photocatalysis\cite{schneider2014understanding, selloni2024aqueous}.
At the TiO$_2$ anatase(101)-water interface, 
DeePMD simulations showed a relatively small extent of water dissociation, $\sim$6 \%\cite{andrade2020free}. Employing umbrella sampling simulations to compute the free energy surface for water dissociation, it was shown that molecularly adsorbed water is significantly more stable than dissociated water with a large free energy barrier.
Water dissociation is therefore only observed on timescales longer than those accessible with AIMD. 
Subsequent DeePMD-based investigation of this system in conjunction with infrared spectroscopy revealed significant restructuring of the hydrogen bonding network as water coverage increased, transitioning from 1D chains at monolayer coverage to 2D and 3D networks at higher coverages\cite{o2023elucidating}. 
Non-equilibrium MD simulations of thermal transport across the TiO$_2$ anatase(101)-water interface using DeePMD have highlighted the important role of water dissociation on the interfacial vibrational density of states (VDOS)\cite{li2023thermal}. Using empirical potentials instead, which do not capture water dissociation, resulted in higher interfacial thermal conductance due to substantial differences in the interfacial VDOS. 
Similar simulations were carried out for the copper-water interface\cite{li2023thermalcu}.

At the TiO$_2$ rutile(110)-water interface, odd-even oscillations in the surface hydroxylation level with respect to the number of O-Ti-O trilayers have been observed in DeePMD studies\cite{zhuang2022resolving,wen2023water}. Similar oscillations were identified in ab initio studies\cite{harris2004molecular}, though computational expense limited these investigations to thin slabs and low water coverage. A PBE+D3-based MLP resulted in an average hydroxylation fraction of 2~\% for thick slabs\cite{zhuang2022resolving}, whereas a SCAN-based MLP resulted in a notably higher value of 22~\%\cite{wen2023water}. These discrepancies likely stem from the different density functionals underlying the respective potentials. 
Water dissociation was found to proceed via two possible mechanisms; direct proton transfer and indirect transfer via a solvent molecule\cite{wen2023water}, which differs from anatase(101), where the direct mechanism was not observed\cite{andrade2020free} due to the larger distance between adjacent surface Ti and O sites.
To paint a more general picture of TiO$_2$-water interfaces, seven different low-index TiO$_2$-water interfaces were jointly investigated with HDNNPs\cite{zeng2023mechanistic}. 
Here, free energy surfaces from metadynamics simulations revealed that water dissociation is thermodynamically favourable on anatase(100), anatase(110), rutile(001), and rutile(011), whereas molecular adsorption is favoured on anatase(101), rutile(100) and rutile(110). 
Surface dependent activity was also observed in DeePMD simulations of the TiS$_2$-water interface, where water dissociation was observed on only one of four investigated surfaces\cite{li2023characterizing}.
Here, free energy profiles from umbrella sampling simulations showed this to be the only surface where water dissociation was thermodynamically and kinetically favourable, which was attributed to the unique presence of both four-fold-coordinated Ti and one-fold-coordinated S surface atoms.

Some oxides feature a particularly complex electronic structure, where metal ions exist in multiple oxidation states. This is the case for LiMn$_2$O$_4$, whose interfaces with water
were investigated using an HDNNP\cite{eckhoff2021insights}.
Here, Mn ions coexist in the Mn$^{\mathrm{IV}}$ and high-spin Mn$^{\mathrm{III}}$ oxidation states, necessitating the use of a hybrid density functional to obtain the correct oxidation state distribution. 
Notably, the valency of the manganese ions has been linked to LiMn$_2$O$_4$'s water oxidation properties\cite{cady_tuning_2015}. 
Even though HDNNPs do not include explicit information on the electronic structure, it was shown that this information is learned implicitly and the oxidation state distribution at the surface could be recovered from the predicted dynamics. The oxidation state distribution and water dynamics at the interface, including water dissociation and proton transfer reactions, were found to differ significantly between the investigated surface orientations and terminations.

MLPs have also been used to investigate solvation dynamics at the hematite(001)-water interface\cite{schienbein2022nanosecond}, proton transfer mechanisms at the water interfaces with GaP(110)\cite{fan2023molecular}, 
as well as ice nucleation on microcline feldspar\cite{piaggi2024first} and the dynamics of surface K$^+$ ions at the muscovite-water interface\cite{raman2024ab, raman2024insights}. 
DeePMD, in conjunction with Deep Wannier models\cite{zhang_deep_2020, sommers_raman_2020} for the prediction of 
atomic dipole moments and polarizabilities, has been employed to investigate the interfacial water structure at the $\alpha$-Al$_2$O$_3$(0001) surface through computational Sum-Frequency Generation (SFG) spectroscopy\cite{du2024revealing}. 
At the water interfaces of IrO$_2$(110)\cite{raman2023acid} and SnO$_2$(110)\cite{jia2024water}, proton transfer mechanisms and acid-base equilibrium properties were investigated, including the calculation of pKa values from calculated free energy differences based on enhanced sampling simulations\cite{raman2023acid} and a counting analysis\cite{jia2024water}. 
Alternatively, thermodynamic integration may be used to compute pKa values. 
Recent studies have demonstrated the utility of MLPs in this context for aqueous molecules\cite{wang2022automated} and transition-metal complexes\cite{wang2024accelerating} with DeePMD, and for BiVO$_4$ in water using a committee of HDNNPs\cite{schienbeindata}.

Theoretical models of the Pt-water interface have been of great interest for a long time due to the frequent use of Pt electrodes in electrocatalysis.
HDNNP-based MD simulations of the Pt(111)-water interface showed the formation of a double layer\cite{mikkelsen2021water}. However, unlike the ordered bilayer that was assumed to form under ultrahigh vacuum conditions, the interfacial water structure was found to be dynamically changing due to repulsive interactions between adsorbed water molecules, leading to a semi-ordered structure. The transfer time of water molecules from the secondary water layer to the surface was found to be around 30~ps, while transfer from the water bulk takes~500 ps, far exceeding the timescales accessible using DFT. When comparing the coverage dependence of hydroxyl adsorption between the conventional bilayer and explicit water model, a different trend was observed at large coverage, where explicit solvent molecules were found to significantly reduce the adsorption barrier\cite{mikkelsen2022structure}. 
Furthermore, the effect of steps at the Pt(211) surface on the properties of the interfacial water has been investigated using DeePMD\cite{wang2024}, revealing distinct physi- and chemisorption patterns and anisotropic dynamics along steps. 
Recent work investigating contact layer water at various surfaces, including Pt, Au, graphene, and MoS$_2$, has highlighted notable differences in the short-range anisotropy and long-range homogeneity of the oxygen-oxygen pair correlation functions of water at these surfaces\cite{gading2024role}.
These variations were found to impact phenomena such as nanofluidic slip and diffusio-osmotic transport, with the in-plane corrugation of the contact layer playing a key role.

\textbf{\textit{Imperfect Surfaces}}
In experiment, surfaces are rarely pristine. 
Capturing the complexity of realistic surfaces in interface systems is challenging, since the precise nature of defects in experiment is often not known. MLP-driven simulations can provide valuable insight which may help to identify the nature of defects and their properties. For instance, recent electrochemical scanning tunnelling microscopy experiments revealed a double-row pattern in the structure of interfacial water at the TiO$_2$ rutile(110)-water interface\cite{sun2024step}. DeePMD simulations were able to quantitatively reproduce the experimental pattern, but only after a [$1\overline{1}1$] step was included in the computational model\cite{sun2024step}. Thus, the simulations point to the presence of [$1\overline{1}1$] steps on this surface. 
Modelling stepped surfaces of this kind requires thousands of atoms, rendering ab initio MD unfeasible. 

To date, only a few MLP-driven studies of solid-liquid interfaces have considered non-pristine surfaces. In relatively early work, the diffusion of surface adatoms and vacancy defects at low-index and stepped Cu-water interfaces was investigated using HDNNPs\cite{kondati2017self}. 
Free energy profiles for adatom and vacancy diffusion from metadynamics simulations were compared to barriers obtained from nudged-elastic-band (NEB) calculations in vacuum, revealing that solvation significantly affects barrier heights for adatom diffusion but has only a marginal impact on vacancy diffusion. 
Water adsorption on MgO and magnetite surfaces, including step and line defects for the latter, has been investigated using VASP's on-the-fly MLPs, based on GAPs\cite{li2020machine}. 
Defect segregation and the impact of oxygen vacancies at zirconium oxyde- and oxynitride-water interfaces have been investigated using MLP-driven Monte Carlo (MC) simulations, showing that water adsorbs preferably at zirconium sites surrounded by vacancies but not on the vacancies themselves\cite{nakanishi2023structural}.

Amorphous systems have shown promise as heterogeneous catalysts, sometimes even outperforming their crystalline counterparts\cite{goldsmith2017beyond}. The amorphous TiO$_2$-water interface has been investigated with DeePMD\cite{ding2023modeling}, showing that interfacial water is far more disordered than observed for crystalline facets, resulting in a $\sim$10 fold increase in the diffusivity of interfacial water molecules compared to the rutile(110)\cite{wen2023water} and anatase(101)-water interfaces\cite{calegari2018structure}. 

Surface reconstruction has been investigated at perovskite oxide-water interfaces using VASP's on-the-fly MLPs\cite{li2024molecular}.
Here, oxygen exchange from the lattice to the liquid, facilitated by the relatively weak Co–O bond, was found to lead to the formation of surface peroxo species and O$_2$ gas, suggesting that Co atoms may act as active sites for the oxygen evolution reaction. 
Recently, the Subsurface Cation Vacancy (SCV) model of the magnetite(001)/water interface, where subsurface cation vacancies stabilize the reconstructed surface\cite{bliem2014subsurface}, has been investigated with HDNNPs\cite{romano2024structure}, uncovering new low-coverage water ground states and anisotropic diffusion of water on the surface. 

\textbf{\textit{Nanoconfined Systems}}
Nanoconfinement has been shown to significantly affect the performance of electrocatalysts\cite{wordsworth2022influence}. 
Understanding the effect of confinement at the atomistic scale can provide valuable insights for optimizing the design and implementation of nanoconfined systems in electrocatalysis.
HDNNPs have been used to investigate water permeation between stacked layers of hexagonal boron nitride (hBN)\cite{ghorbanfekr2020insights}, revealing a significant increase in the water self-diffusion coefficient under strong confinement—a result not captured by simpler water models such as SPC/E or TIP4P due to a poor description of water-surface interactions.
Studies of nanoconfined water between graphene sheets using DeePMD\cite{zhao2022deep, liu2023transferability} have shown an increase in water permeability under confinement.
From experiment, it is known that water transport is significantly faster in graphene nanotubes compared to hBN. However, the origin of this behavior was not well understood\cite{secchi2016massive}. 
HDNNP-based investigation of these systems could reproduce the experimental trend, and provide atomic-scale insight into the different water transport properties of these systems\cite{thiemann2022water}. Hydrogen atoms from water were found to interact strongly with nitrogen atoms at the hBN surface, leading to substantial residence time and slow transport. Such interactions are not present for graphene nanotubes, leading to faster transport dynamics. 
Finally, DeePMD investigation of nanoconfined water between TiO$_2$ anatase(101) slabs showed the formation of 1D water chains on the surface, leading to anisotropic surface hydrogen diffusion, as well as significantly reduced surface hydroxyl lifetimes\cite{kwon2024confinement}.

\textbf{\textit{Beyond Pure Water}}
Given the complexity of structural motifs and reactivity of even pure water at a solid surface, few studies have considered additional species at solid-water interfaces so far. Including reactants in simulations of solid-liquid interfaces is, of course, crucial for extending these studies to real catalytic processes, making this an important area for future research. 

The effect of explicit solvents on adsorbate properties at  the Cu(111)-water interface has been investigated with MLPs\cite{chen2023accelerating}. Binding energies of CO and OH obtained using implicit vs. explicit solvent models showed significant discrepancy, particularly for OH, which actively participates in the hydrogen bonding network. Metadynamics simulations were used to compute the free energy barriers for C–H bond breaking of ethylene glycol over Cu(111) and Pd(111), revealing distinct reaction pathways on the two surfaces, with bond breaking occurring more readily on Pd(111).

The oxygen reduction reaction (ORR) at the Au(100)-water interface has been investigated with metadynamics simulations\cite{yang2023neural} employing the message-passing graph neural network PaiNN\cite{schutt2021equivariant}. An O$_2$ molecule introduced into the liquid phase was found to readily migrate to the surface. The ORR was shown to proceed via an associative reaction pathway with a low barrier of 0.3~eV, consistent with experimental observations of high ORR activity on Au(100), but showing slight mechanistic differences from the pathway proposed in earlier theoretical work\cite{norskov2004origin}. 

Protonated water confined between MXene sheets, a class of 2D transition metal carbides or nitrides with applications in (photo)catalysis\cite{kuang2020mxene}, has been studied using DeePMD simulations\cite{houproton}. The presence of Eigen and Zundel cations was found to influence the orientation of nearby water molecules, which in turn inhibited water-induced oxidation of the surface, a process previously identified in DeePMD simulations of the MXene interface with pure water\cite{hou2023unraveling}. Additionally, an unusual hexagonal ice phase, stabilized by hydrogen bonds to the MXene surface, was observed for low proton concentrations at room temperature.

Moreover, the behavior of carboxylic acids at the TiO$_2$ anatase(101)-water interface has been investigated with DeePMD simulations\cite{raman2024long}. At high acid coverage, a transition from bidentate to monodentate adsorption was observed, accompanied by the co-adsorption of a water molecule on the vacated Ti$_{5c}$ site. This configuration was found to be stable over long timescales, which could not be observed in shorter AIMD simulations.
Furthermore, the effect of varying the pH of anatase(101)-electrolyte solutions has been investigated with DeePMD simulations\cite{zhang2024electrical} 
Here, a Deep Wannier model was trained to reproduce DFT Wannier centers, enabling the calculation of the electrostatic potential profile and interfacial capacitance. The results highlight the complexity of the ion distribution at solid-electrolyte interfaces which is not captured by mean-field theories such as the Gouy-Chapman-Stern (GCS) model\cite{bard2022electrochemical}. For instance, positively charged counterions were found to approach the surface more closely than negatively charged counterions due to the electronegative surface oxygen atoms, regardless of the pH and surface charge. Consequently, the interfacial capacitance was found to be larger for basic solutions with negative surface charge compared to acidic solutions with positive surface charge, consistent with experiment\cite{be1968adsorption} but in contrast to the GCS prediction. 

The additional complexity introduced when adsorbates are included at solid-liquid interfaces presents a challenge, requiring more exhaustive training to fully capture the configuration space. 
To tackle this, a hybrid QM, FF, and MLP approach was recently proposed for computing aqueous-phase adsorption free energies. Here, an MLP is used only for water-surface interactions, while water-water and adsorbate-surface interactions are treated at different levels of theory\cite{zare2024hybrid}. This approach retains some of the speedup advantages of MLPs, while simplifying the active learning problem since the MLP needs only to cover a subset of the full configuration space. \\

\section{Discussions, Challenges, and Outlook}\label{sec:discussion}

The broad application of MLPs discussed in the previous section demonstrates that MLPs have matured beyond their initial proof-of-concept phase to the point where the investigation of realistic catalytic systems is now possible.
From metal clusters and solid surfaces to gas-surface dynamics and solid-liquid interfaces, MLPs are becoming ubiquitous tools in the theoretical study of catalytic processes.
However, the construction and application of MLPs, particularly in heterogeneous catalysis, requires great care, since ML models are, to some extent or entirely,
agnostic to the underlying physics of the system, learning the shape of the potential energy surface from the data provided. This flexibility, while offering a significant advantage over empirical potentials, demands caution, i.e., they are still far from being a black-box solution. In the following, several key considerations concerning the use of MLPs in heterogeneous catalysis are discussed.

\textbf{\textit{Reference Configurations and Data Efficiency}}
To train MLPs, reference configurations along with their associated energies, forces, and possibly charges, spins and stresses, must be provided. However, selecting the most relevant configurations is nontrivial, as they should ideally encompass the atomic environments likely to be encountered in the target simulations. Predicting all such relevant environments beforehand is practically impossible. Including as many relevant structures as possible improves the accuracy and reliability of the MLPs. However, this raises concerns about data efficiency. Due to the computational cost of reference electronic structure calculations and the demands of training on large datasets, minimizing the amount of data required to build the MLP is crucial. 
This is particularly important for catalytic systems, where the training data must not only include bulk solid and liquid phases but also complex interfaces (see Figure~\ref{fig:MLP-construction}).

These challenges may be addressed with active learning schemes~\cite{P6548,P3114,schran2020committee,vandermause2022active, perego2024data}.
Active learning is an iterative process that enhances MLP accuracy by selectively improving underrepresented parts of the configuration space. One common approach is query by committee (QbC)\cite{seung1992query, krogh1994neural, smith2018less, schran2020committee}, where the variance in energy (or force) predictions across an ensemble (or ``committe'') of similarly-trained models is used as a measure of uncertainty. Structures with large committee variance, which are thus underrepresented in the training set, can be identified as important data points to include in further training cycles.
A partially trained committee model may be used to drive MD simulations at the intended production conditions. In this way, the relevant configuration space
is explored autonomously. 
When high-uncertainty configurations are encountered, additional reference calculations can be performed and added to the dataset, followed by retraining the MLP. This process can be repeated over multiple iterations, 
until the configuration space of interest is sufficiently well sampled. 

Schemes to identify reaction coordinates and improve the sampling of rare transitions in complex molecular systems have also been proposed~\cite{yang2022using,jung2023machine,PNAS_Tiwary_2023}.
Another approach to enhance data efficiency 
involves using equivariant descriptors, 
which have been shown to significantly reduce the amount of training data required~\cite{batzner20223}. This approach is part of a broader design philosophy that advocates \textit{exploiting} the symmetry operations in 3D space, rather than merely \textit{respecting} them~\cite{batzner2023advancing,thomas2018tensor,weiler20183d}.

\textbf{\textit{Transferability}} The concept of \textit{transferability} is used with varying meanings and in different contexts in the literature. In the context of empirical potentials, which are inherently fitted to a limited number of experimental or high-quality quantum mechanical data, transferability usually refers to the potential's ability to reproduce properties of systems not included during its construction. This issue is somewhat mitigated in MLPs since they are trained on the potential energy surface itself rather than specific properties or parameters. 
However, MLPs perform poorly when applied outside the configuration space covered during training.

Active learning, discussed in the preceding section, can be used to systematically extend the model to accurately cover new thermodynamic (or compositional) state points, however this requires some effort if many such state points are of interest.
Some recent graph neural networks have shown impressive extrapolation abilities beyond the training dataset, however such claims must be approached with caution, and further evaluations are needed to verify their reliability. 
Another promising approach is to utilize transfer learning and foundation models as initial frameworks for MLP development, which could potentially improve transferability and out-of-distribution performance \cite{batatia2023foundation,falk2024transfer}.

\textbf{\textit{Reference Electronic Structure Method}} Although MLPs are not restricted to any specific electronic structure technique, DFT, particularly at the GGA level of theory, remains the workhorse for heterogeneous catalysis. This is because the complexity of catalytic interfaces typically requires thousands of electronic structure calculations on systems large enough to include the bulk catalyst, solvents, and their interfaces in the training dataset. 
However, GGA functionals suffer from several limitations, such as self-interaction error, resulting in underestimated band gaps, as well as inaccurate prediction of reaction barriers and structural properties. These shortcomings are particularly pronounced in systems with localized or strongly correlated electrons such as transition metal oxides.
Moving beyond GGA functionals can help mitigate these issues by incorporating additional information, such as kinetic energy density in meta-GGAs or exact exchange from Hartree-Fock theory in hybrid functionals, leading to more accurate and reliable predictions. Additionally, corrections for missing dispersion interactions are often required, especially for systems including water\cite{morawietz2016van}. 
While beyond-DFT methods have already been employed to construct MLPs for bulk systems, for example using coupled-cluster theory for liquid water\cite{P6321,P6362,P6360} and the random-phase approximation (RPA) for zirconium oxide\cite{liu2022phase},
it seems unlikely that these methods will become mainstream for studying catalysis in the near future for most practical applications. However, this trend may change with increasing efficiency of electronic structure calculations.

The choice of DFT functional for constructing the training dataset must be considered carefully. A well-established procedure for verifying the suitability of a given functional is to benchmark the description of experimentally accessible observables e.g., slab lattice parameters, solvent density, or surface adsorption energies in the case of solid-gas interfaces, against experimental results or data from higher levels of theory. Multi-component systems generally pose an increased challenge, as some GGA functionals are known to better describe solvents, while others might be better suited to describe solids. Transition metal oxide systems~\cite{eckhoff2020hybrid}, for example, are often well described by PBE-based hybrid functionals such as HSE06 or PBE0, but their weak description of water is well-documented.\cite{gillan_perspective_2016} The inefficacy of DFT, particularly GGA functionals, in describing the adsorption of molecules on catalysts is also well-known. However, a detailed discussion of the advantages and disadvantages of various functionals and electronic structure techniques is beyond the scope of this article. For further information, the reader is referred to more detailed reports on the subject (see Ref.~\citenum{sauer2024future}). To conclude, in practice, when generating the reference data for a catalytic system the  electronic structure method that can describe the overall system at the desired accuracy has to be chosen with great care while the method also still needs to be efficient enough to enable a sufficient sampling of the relevant configuration space.

\textbf{\textit{Accuracy}} 
Comparing the accuracy of different MLP models is a complex task. For instance, while MLP A may outperform MLP B on a specific dataset, it could underperform on another, making it challenging to find a balanced comparison between MLPs in the literature, though some initial attempts have been made\cite{stark2024benchmarking,P6531,chen2023accelerating}. The current practice for evaluating new MLP models involves comparing their energy and force root mean square errors (RMSE) on benchmark datasets. However, it has been shown that although such evaluations are necessary, they are not sufficient~\cite{fu2022forces,P6302}. It is important to note that most state-of-the-art MLP implementations generally achieve energy errors around 1~meV/atom and force errors on the order of 100~meV/\AA{}. These error magnitudes are significantly smaller than the uncertainties introduced by factors like the choice of exchange-correlation functional in DFT. Furthermore, studies have demonstrated that the discrepancy between DFT and experimental observables is greater than that between MLPs and DFT~\cite{montero2024comparing,omranpour2024high}. Since MLPs are inherently limited by the accuracy of the underlying electronic structure methods, a practical recommendation when applying MLP models to heterogeneous catalysis is to prioritize generating high-quality reference data using the most accurate electronic structure methods and ensuring convergence, rather than over-investing on the selection of a particular MLP.

\textbf{\textit{Long-Range Interactions}}
Most MLPs are local, i.e., they construct the potential energy surface as a sum of environment dependent atomic energy contributions. However, some interactions, such as dispersion or electrostatics, can reach far beyond the typical cutoff range of 5 to 10 \AA{}. During training, the tail of these interactions is not truncated but partitioned into the local atomic environments. For most condensed phase systems this approach is sufficient because these interactions are effectively screened and therefore, the contribution beyond the cutoff is very small. However, when they are still significant, they act like noise during training, which can reduce the quality of the fit.\cite{yue_when_2021} In condensed phase systems containing e.g., electrolytes or ionic liquids, or solid-gas interface systems, where screening effects are significantly reduced, an explicit treatment of long-range interactions can therefore become necessary to achieve the desired accuracy.\cite{yue_when_2021}\\
MLPs of the third and fourth generation have been extended to include long-range electrostatics by learning environment dependent charges from which electrostatic energies and forces can be computed. Similarly, long-range dispersion corrections such as the Tkatschenko-Scheffler\cite{tkatchenko_accurate_2009} and exchange-hole dipole moment\cite{becke_exchange-hole_2005, becke_exchange-hole_2007} model have been combined with MLPs by learning environment dependent Hirshfeld volumes,\cite{muhli_machine_2021} and Hirshfeld volumes and exchange-hole moments~\cite{tu_neural_2023}. Additionally, long-range dispersion can also be included without falling back on an additional ML model by applying Grimme's geometry dependent DFT-D3 correction~\cite{grimme_consistent_2010}.

\textbf{\textit{Non-Local Interactions}}
For many systems, accounting only for local interactions is sufficient to achieve an accurate description.\cite{unke_spookynet_2021} However, for some systems, accounting for global interactions becomes necessary to predict the correct dynamics. Ko et al.\cite{ko_fourth-generation_2021} demonstrated that using a model containing an Au$_2$ cluster supported on a MgO substrate. On pristine MgO, the Au$_2$ cluster takes an upright configuration where only a single Au atom is adsorbed to the surface. Introducing Al doping, which replaces some Mg sites below the surface, leads to a global charge redistribution and different optimal Au$_2$ cluster configuration that is parallel to the surface where both Au atoms directly interact with the surface. In this model the doping is introduced at a distance beyond the cutoff. Consequently, local MLPs are unable to predict the change in adsorption geometry of the Au$_2$ cluster when introducing the dopant. Using a 4G-HDNNP, Ko et al. are able to predict the correct adsorption geometry. Unke et al.\cite{unke_spookynet_2021} also demonstrated that their proposed MPNN architecture SpookyNet, that includes nonlocal interactions, is able to describe this model accurately.

\textbf{\textit{Electrode Potentials and External Electric Fields}}
In a realistic electrochemical setting, reactions normally occur under a constant electrode potential. This requires free exchange of electrons with a reservoir, posing a significant challenge for molecular dynamics-based atomistic simulations that typically sample ensembles which conserve charge instead\cite{nielsen2015towards, levell2024emerging}. Several approaches for including constant electrode potentials in MD simulations have been proposed\cite{bonnet2012first, melander2019grand, dufils2019simulating, deissenbeck2021dielectric}, however this field is at a relatively early stage of development. 

An important first step towards more realistic modelling of electrochemical systems is the ability to apply external electric fields in simulations. Electric fields may significantly influence catalytic activity, altering reaction pathways and energetics by modulating charge distributions and bonding character\cite{che2018elucidating, leonard2021electric}. Several distinct methods to incorporate electric fields in MLP-driven MD simulations have been reported\cite{christensen2019operators, gastegger2021machine, shao2022finite, gao2022self, zhang2023universal, joll2024machine}. 
So far, applications of these methods have been limited to relatively simple systems such as liquid water\cite{gao2022self, joll2024machine} and molecules in vacuum or solution\cite{christensen2019operators, gastegger2021machine, shao2022finite, zhang2023universal}. Taking advantage of these ML-accelerated methods to simulate catalytic reactions under electric fields over large time and length scales, particularly in complex heterogeneous environments, is an interesting area of research for the near future.

\textbf{\textit{Nonadiabatic Effects and Multiple States}}
MD simulations are most commonly performed under the Born-Oppenheimer approximation, which separates electronic and nuclear degrees of freedom due to their large difference in mass. In this framework, the electronic wavefunction is assumed to adapt instantaneously to changes in nuclear positions such that nuclei evolve on a single adiabatic PES.
Nonadiabatic effects, which require a description beyond the Born-Oppenheimer approximation, play an important role in processes such as proton-coupled electron transfer reactions that govern many catalytic reactions\cite{weinberg2012proton, solis2014proton, hutchison2024nonadiabatic}, as well as in the dynamics of photoexcited charge carriers in photocatalysis\cite{akimov2013nonadiabatic, you2021nonadiabatic, you2024correlated}. 
Accurate modelling of excited state dynamics often requires computationally expensive methods for the calculation of excited states, drastically limiting the accessible time and length scales that can be probed in simulations of nonadiabatic dynamics, for example using mixed quantum-classical dynamics approaches\cite{crespo2018recent}. Thus, machine learning acceleration for excited state dynamics could offer the same speed-up benefits that have revolutionized the field of ground-state molecular dynamics\cite{westermayr2020machine}. 

Early work employed low-dimensional NNPs to represent constrained DFT (CDFT) spin states of O$_2$ on a frozen Al(111) slab \cite{P1388, P1820, P1996}, enabling investigation of nonadiabatic spin transitions through fewest-switches surface hopping simulations\cite{P2536}. 
More recently, several studies have demonstrated the potential of MLPs in excited state dynamics simulations by utilising machine-learned adiabatic excited states\cite{hu2018inclusion, chen2018deep, dral2018nonadiabatic, westermayr2019machine, westermayr2020combining, li2021automatic, axelrod2022excited}, or diabatic states from CDFT\cite{meng2024first}. However, challenges remain due to the complexity of representing multiple excited states, especially if a large number of states is required, as well as difficulties in accurately reproducing the nonadiabatic coupling in strong-coupling regions where this quantity exhibits sharp peaks\cite{dral2018nonadiabatic}.

We note that MLPs, as originally introduced, are designed to represent a single potential energy surface by constructing a structure-energy relationship.
To represent multiple electronic states, one may either use individual MLPs for each state or employ extended architectures, such as models that predict multiple states simultaneously or encode each state as input to a single model\cite{westermayr2020neural}.
The ability to describe multiple states with MLPs is also critical in the context of magnetic materials, where the ground state magnetic order may change after the introduction of defects or dopants\cite{eckhoff_high-dimensional_2021}, a topic of high interest in catalysis. 
Recently, adapted spin-dependent atom-centered symmetry functions have been proposed for this purpose\cite{eckhoff_high-dimensional_2021}, as well as MPNN architectures that embed electronic structure information into their representation\cite{unke_spookynet_2021, yuan_equivariant_2024}.

\textbf{\textit{Nuclear Quantum Effects}} 
In addition to the Born-Oppenheimer approximation discussed above, a second approximation is often employed in MD simulations: the classical treatment of atomic nuclei as point particles which evolve in phase space according to Newton’s laws. 
Nuclear quantum effects (NQEs), such as zero-point energy and tunnelling, are particularly important at low temperature and in systems involving light elements, though they can significantly influence properties even at room temperature\cite{bocus2023nuclear}. Path integral molecular dynamics, and related methods like ring polymer molecular dynamics, have emerged as efficient approaches to incorporate NQEs into atomistic simulations\cite{markland2018nuclear}. These methods involve sampling an extended classical phase space by representing each physical atom by a set of “beads”, typically around 10-100\cite{althorpe2021path}. Calculations can be efficiently parallelized over beads, resulting in a computational overhead of only a factor of 10-100, far smaller than the gains achieved by replacing DFT with MLPs. Several studies have demonstrated the use of machine learning potentials to accelerate path-integral MD simulations\cite{kapil2016high, cheng2016nuclear, hellstroem2018nuclear, schran2018converged, cheng2019ab, schran2019quantum, schran2019automated, kapil2020inexpensive, yao2020temperature, yao2021nuclear, schran2021transferability, li2022using, kimizuka2022artificial, liu2023mechanistic, ple2023routine, bocus2023nuclear, kwon2023accurate, kapil2022first, lin2023temperature, atsango2023developing, P6360}.
Of particular interest to heterogeneous catalysis is a recent study on proton hopping kinetics in zeolites\cite{bocus2023nuclear}. NQEs, accounted for using MLP-accelerated ring polymer molecular dynamics, were found to drastically reduce activation energies for proton hopping and increase hopping rates. At 273 K, the proton hopping rate increased by a factor of 65, and even at 473 K it remained 7 times larger than in classical simulations. 
It can be anticipated that accounting for NQEs in atomistic simulations of catalytic systems will become considerably more common with the increasing prevalence of machine learning potentials. 

\section{Conclusions}\label{sec:conclusions}

The knowledge-based investigations of heterogeneous catalysis have traditionally been challenged by three well-known gaps: the complexity gap, the materials gap, and the pressure gap~\cite{schlogl2015heterogeneous, mou2023bridging}. One major example of the complexity gap that has slowed the theoretical modeling of heterogeneous catalysis is the gap between the time and length scales accessible with a quantum-mechanical treatment of atomic interactions and those required for realistic modeling of complex catalytic interfaces. Recently, Machine Learning Potentials have emerged as a powerful solution to bridge this \textit{modeling gap}. Their ability to accurately learn reactive atomic interactions has allowed researchers to move beyond traditional tools and explore catalysis under more realistic conditions. With carefully constructed MLPs, simulations involving thousands of atoms over nanosecond timescales can now achieve quantum-mechanical accuracy, overcoming the limitations of conventional methods such as \textit{ab initio} molecular dynamics and classical force fields.

In this article, we have explored the role of MLP-driven atomistic simulations in advancing our understanding of catalytic systems, including gas-phase and liquid-phase interfaces. Insights gained include elucidating reaction mechanisms, assessing solvent effects, and investigating the influence of defects and interfaces on catalytic processes. 
Furthermore, MLPs enable simulations under realistic \textit{operando} conditions, capturing essential temperature effects and dynamic behaviors crucial for interpreting catalytic activity in real-world applications. 
These results further highlight the potential of MLPs to advance our understanding of complex phenomena such as proton transfer, surface reconstructions, and nanoconfinement.

Despite their advantages, MLPs still require careful selection of training data, electronic structure methods, and validation of their transferability across various conditions. With the increasing availability of advanced yet user-friendly MLP packages, we expect the adoption of MLPs to become increasingly prevalent in heterogeneous catalysis research.
Consequently, a shift from the traditional static view of catalytic systems to one that explicitly accounts for dynamic effects is expected. Such progress is likely to deepen our understanding of catalytic processes, ultimately guiding the design and optimization of new catalysts. Addressing current challenges, including improving data efficiency, capturing long-range interactions, and enhancing model robustness, is key to the further progress and broader adoption of MLPs in catalysis science.

\begin{acknowledgments}
AO thanks the Deutsche Forschungsgemeinschaft (DFG) for funding in TRR/CRC 247 (A10, project number 388390466), JE thanks the DFG for funding in CRC 1633 (C04, project number 510228793), and KNL thanks the DFG for funding in CRC 1073 (C03, project number 217133147). Moreover, we are grateful for support by the DFG under Germany's Excellence Strategy – EXC 2033 RESOLV (project-ID 390677874).
\end{acknowledgments}

\bibliography{literature}

\bibliographystyle{achemso}

\end{document}